\documentclass[a4paper,twocolumn,showpacs,superscriptaddress,prl]{revtex4}
\usepackage[english]{babel}
\usepackage{graphicx}
\usepackage{xspace}
\usepackage{mathptmx}

\begin{document}

\title{Sub-unit cell layer-by-layer growth of Fe$_3$O$_4$, MgO, and Sr$_2$RuO$_4$ thin films}

\author{D.~Reisinger}
\email{Daniel.Reisinger@wmi.badw.de}
\affiliation{Walther-Mei{\ss}ner-Institut, Bayerische Akademie der
Wissenschaften, Walther-Mei{\ss}ner Str.~8, 85748 Garching, Germany}

\author{B.~Blass}
\affiliation{Walther-Mei{\ss}ner-Institut, Bayerische Akademie der
Wissenschaften, Walther-Mei{\ss}ner Str.~8, 85748 Garching, Germany}

\author{J.~Klein}
\affiliation{II. Physikalisches Institut, Universit\"{a}t zu K\"{o}ln,
Z\"{u}lpicher Str.~77, 50937 K\"{o}ln, Germany}

\author{J.~B.~Philipp}
\affiliation{Walther-Mei{\ss}ner-Institut, Bayerische Akademie der
Wissenschaften, Walther-Mei{\ss}ner Str.~8, 85748 Garching, Germany}

\author{M.~Schonecke}
\affiliation{Walther-Mei{\ss}ner-Institut, Bayerische Akademie der
Wissenschaften, Walther-Mei{\ss}ner Str.~8, 85748 Garching, Germany}

\author{A.~Erb}
\affiliation{Walther-Mei{\ss}ner-Institut, Bayerische Akademie der
Wissenschaften, Walther-Mei{\ss}ner Str.~8, 85748 Garching, Germany}

\author{L.~Alff}
\email{Lambert.Alff@wmi.badw.de}
\affiliation{Walther-Mei{\ss}ner-Institut, Bayerische Akademie der
Wissenschaften, Walther-Mei{\ss}ner Str.~8, 85748 Garching, Germany}

\author{R.~Gross}
\affiliation{Walther-Mei{\ss}ner-Institut, Bayerische Akademie der
Wissenschaften, Walther-Mei{\ss}ner Str.~8, 85748 Garching, Germany}

\date{received August 23, 2002}




\begin{abstract}
The use of oxide materials in oxide electronics requires their controlled
epitaxial growth. Recently, it was shown that Reflection High Energy
Electron Diffraction (RHEED) allows to monitor the growth of oxide thin
films even at high oxygen pressure. Here, we report the {\em sub\/}-unit
cell molecular or block layer growth of the oxide materials Sr$_2$RuO$_4$,
MgO, and magnetite using Pulsed Laser Deposition (PLD) from stoichiometric
targets. Whereas for perovskites such as SrTiO$_3$ or doped LaMnO$_3$ a
{\em single} RHEED intensity oscillation is found to correspond to the
growth of a {\em single} unit cell, in materials where the unit cell is
composed of several molecular layers or blocks with identical
stoichiometry, a sub-unit cell molecular or block layer growth is
established resulting in {\em several} RHEED intensity oscillations during
the growth of a {\em single} unit-cell.
\end{abstract}
\maketitle


The physical properties of thin films are strongly influenced by their
microstructure and morphology which, in turn, are determined by the
deposition conditions and the growth mode. RHEED has proven to be a useful
surface sensitive tool for monitoring {\em in situ} the growth of
semiconductor thin films \cite{Lagally:93}. RHEED has also been
successfully used for studying the growth of the complex copper oxide
superconductors under high vacuum conditions
\cite{Terashima:90,Karl:92,Bozovic:95}. Recently, high pressure RHEED
systems have been developed allowing for the analysis of oxide thin films
grown by Pulsed Laser Deposition (PLD) at high oxygen partial pressure of
up to several 10\,Pa \cite{Rijnders:97,Klein:99}. A common way of using
RHEED in the analysis of thin film growth is the observation of the
electron diffraction pattern during epitaxial growth. In this case
intensity oscillations of the diffraction spots are associated with a
layer-by-layer or Frank-van~der~Merwe growth mode \cite{Terashima:90}. The
deposited material nucleates on the substrate surface forming two
dimensional islands which are coalescing with increasing coverage. This
process results in a periodic roughening and flattening of the film surface
translating into RHEED intensity oscillations.

In this Letter, we report on the study of RHEED intensity oscillations
during the PLD growth of the oxides materials Sr$_2$RuO$_4$, MgO, and
magnetite (Fe$_3$O$_4$) from stoichiometric targets. We have used an ultra
high vacuum Laser Molecular Beam Epitaxy (L-MBE) system with in-situ high
pressure RHEED and a 248\,nm KrF excimer laser \cite{Gross:2000a}. It is
well known that due to the ionic bond character in oxides the different
atomic layers in general will not be charge neutral and therefore the
energetically most favorable growth unit often is a molecular layer which
is composed of one or several atomic layers to obtain charge neutrality.
Therefore, molecular layer or block layer epitaxy\cite{Gross:2000a} is
established for most oxides, whereas atomic layer epitaxy is common for
semiconductor superlattice growth. Our results clearly show that under
suitable deposition conditions a molecular or block layer growth is indeed
established for MgO, Sr$_2$RuO$_4$ and Fe$_3$O$_4$. However, in the
present study the observed thickness of the charge neutral molecular
layers is exceeding that of the minimum charge neutral building blocks of
the respective material, since stoichiometric targets have been used in
the PLD process. In this case, the minimum observed building blocks are
those satisfying both the requirement of charge neutrality and the
chemical composition supplied from the stoichiometric target material. Let
us illustrate this for the example of SrTiO$_3$. This material can be
grown both in alternating charge neutral SrO and TiO$_2$ molecular layers
supplied from SrO and TiO$_2$ targets, and in SrTiO$_3$ layers provided
from a stoichiometric target. Note that in the first case every molecular
layer will produce a RHEED intensity oscillation resulting in two RHEED
oscillations per unit cell. In contrast, in the latter case only a single
oscillation per unit cell is observed, since SrO and TiO$_2$ are supplied
simultaneously and the minimum charge neutral block compatible with the
supplied stoichiometry consists of the whole unit cell. We observed an
equivalent behavior for the growth of MgO, Sr$_2$RuO$_4$, and Fe$_3$O$_4$
using stoichiometric targets. Due to the complex unit cells of MgO and
Sr$_2$RuO$_4$ consisting of identical stoichiometric layers shifted by
lattice translations two RHEED oscillations per unit cell are observed.
For Fe$_3$O$_4$, the unit cells consists of even 4 stoichiometric
sub-units resulting in 4 RHEED oscillations per unit cell.

\begin{figure}[tb]
\centering{%
\includegraphics [width=0.99\columnwidth,clip=]{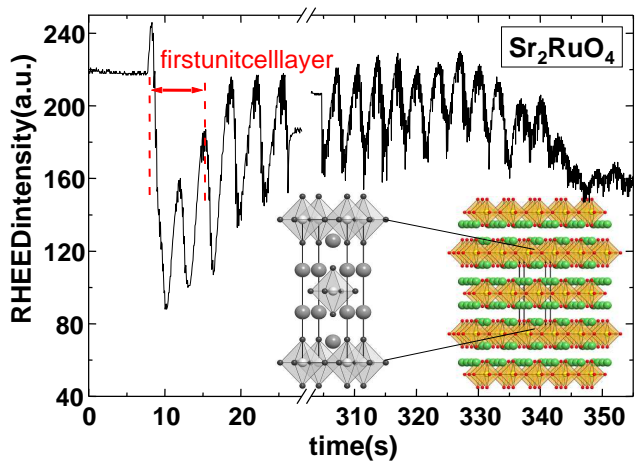}}
 \caption{\small
RHEED intensity oscillations observed during the growth of Sr$_2$RuO$_4$.
During the growth of a {\em single} unit cell {\em two} intensity
oscillations are observed. The inset shows the crystal structure with two
identical stoichiometric building blocks translated by
$(\mathbf{a}+\mathbf{b})/2$ within the $ab$-plane.}
 \label{srruo}
\end{figure}

We first discuss the growth of the unconventional superconductor
Sr$_2$RuO$_4$. This material is isostructural to La$_2$CuO$_4$, which
already has been studied by RHEED \cite{Terashima:90}. For structural
reasons, a similar growth is expected for both materials. Fig.~\ref{srruo}
shows the RHEED intensity oscillations of the $(0,0)$ diffraction spot
recorded during the PLD growth of a $c$-axis oriented Sr$_2$RuO$_4$ film
on a NdGaO$_3$ substrate. The layer-by-layer growth mode is achieved for a
substrate temperature $T_S= 950^\circ$C, an oxygen pressure of
$p_{O_2}=20$\,mTorr, a laser repetition rate of $f_L=5$\,Hz, and a laser
energy density on the target of $P_L=1.2$\,J/cm$^2$. RHEED was performed
with 15\,keV electrons at an incident angle of $2.2^\circ$. To obtain the
number of RHEED oscillations per unit cell, the film thickness has been
determined precisely by X-ray reflectometry and then divided by the number
of observed RHEED oscillations. The derived block thickness corresponds to
half a unit cell ($c/2=6.39\,${\AA}). Considering the crystal structure shown
in the inset of Fig.~\ref{srruo}, this result is intuitive: The unit cell
consists of two identical layers which are displaced within the $ab$-plane
by $(\mathbf{a}+\mathbf{b})/2$, where $\mathbf{a}$ and $\mathbf{b}$ are
the in-plane lattice vectors. Note that after the third oscillation the
intensity of the maxima remains about constant. The growth can be stopped
and continued with almost unchanged modulation depth. After a certain film
thickness the oscillations start to flatten out with an overall decrease
of reflected intensity due to surface roughening because of a transition
to island growth. The layer-by-layer growth mode can be re-established by
thermal annealing. An interesting feature is that the RHEED intensity
starts to {\em increase} immediately after starting the PLD process.
However, this effect is only present for starting the growth on the
substrate, but is absent for starting on the surface of an already grown
Sr$_2$RuO$_4$ film. Therefore, we believe that this effect is not related
to the diffraction conditions, but more likely to a change of the surface
structure and morphology of the substrate due to the partial coverage with
the thin film material. A similar effect has been observed for
isostructural La$_2$CuO$_4$ \cite{Terashima:90} as well as for GaAs
\cite{Dobson:87}. We finally note, that our high quality Sr$_2$RuO$_4$
films have lower resistivity values ($\rho\lesssim 20\,\mu\Omega$cm at
300\,mK) than those reported so far in literature \cite{Schlom:98}.
However, these values are still about ten times larger than those required
for superconductivity ($\rho \leq 1.5\,\mu\Omega$cm).

\begin{figure}[tb]
\centering{%
\includegraphics [width=0.99\columnwidth,clip=]{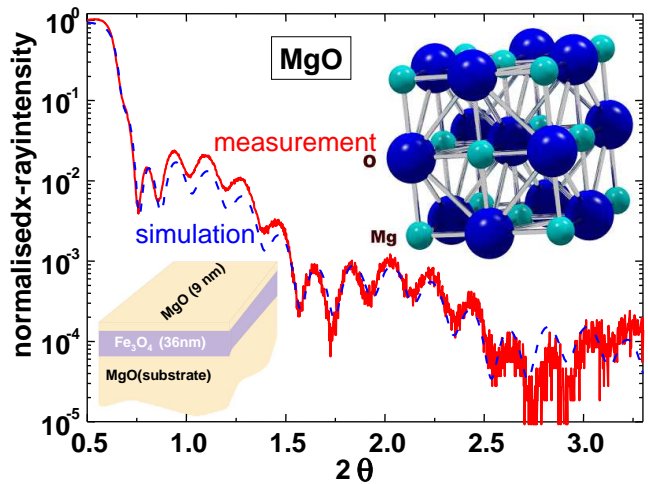}}
 \caption{\small
X-ray reflectometry data (solid line) and simulation data \cite{refsim}
(broken line). The inset shows the MgO unit cell ($a=b=c=0.4211$~nm)
consisting of 2 Mg$^{2+}$ and 2 O$^{2-}$ ions. }
 \label{MgO}
\end{figure}

As the second example we discuss the growth of a MgO cap layer on a
magnetite thin film, which will be discussed below. Both films were grown
under the same growth conditions ($T_S=330^\circ$C, $p_{\rm Ar}=2\times
10^{-3}$\,mbar, $f_L=2$\,Hz, and $P_L \simeq 5$\,J/cm$^2$) leading for both
materials to a layer-by-layer growth mode. First the magnetite thin film
was grown on a MgO substrate with atomically flat surface (3\,{\AA} rms
roughness) followed by the MgO cap layer. Fig.~\ref{MgO} shows X-ray
reflectometry data of a Fe$_3$O$_4$~(36\,nm)/MgO~(9\,nm) bilayer. Fitting
the data\cite{refsim} gives the thickness of the MgO and Fe$_3$O$_4$ layers
with an error of less than $\pm 5$\%. As already described above, from the
MgO layer thickness and the observed number of RHEED oscillations we obtain
the result that two RHEED oscillations correspond to the growth of a single
MgO unit cell shown in Fig.~\ref{MgO}. Again, the intuitive explanation for
this observation is the fact that the unit cell of MgO consists of two
identical stoichiometric layers displaced along the $ab$-plane by
$(\mathbf{a}+\mathbf{b})/2$. After the deposition of the MgO cap layer, the
surface morphology was examined {\em in-situ} by Atomic Force Microscopy.
The rms surface roughness averaged over an area of 1\,$\mu$m$^2$ was below
2\,{\AA}, i.e. even below that of the MgO substrate. This result is consistent
with a dominant layer-by-layer growth mode observed for the whole bilayer.

\begin{figure}[tbh]
\centering{%
\includegraphics [width=0.99\columnwidth,clip=]{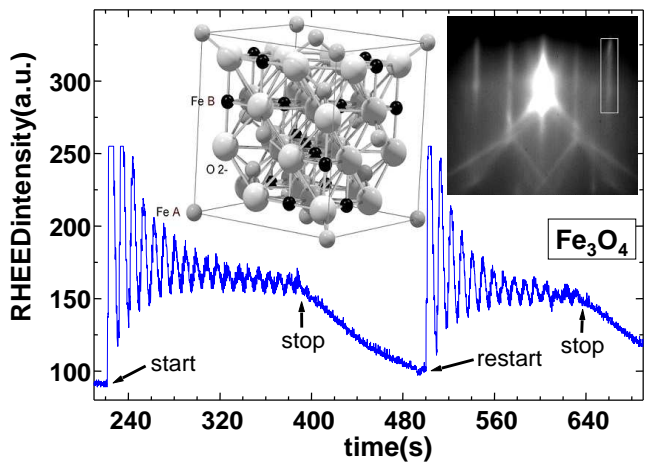}}
 \caption{\small
RHEED intensity oscillations of two series of laser pulses during the
growth of a Fe$_3$O$_4$ thin film on a MgO substrate. The right-hand inset
shows the RHEED pattern. For the RHEED oscillations the intensity is
integrated within the rectangle around the $(0,2)$ spot. The left-hand
inset shows the unit cell of magnetite consisting of four Fe$_{3}$O$_{4}$
formula units.}
 \label{fe3o4}
\end{figure}

As the last example we discuss the growth of Fe$_3$O$_4$ thin films on MgO
substrates. Fig.~\ref{fe3o4} shows the RHEED intensity vs. time which has
several interesting features. First, the modulation depth decreases
continuously with increasing film thickness, if the ablation is not
interrupted. Since the intensity of the maxima {\em de\/}creases while the
intensity of the minima {\em in\/}creases, we ascribe this to a gradual
transition to the step-flow-growth mode which leaves the reflected
intensity unchanged. The layer-by-layer growth mode can be re-established
by annealing the sample for about 100\,s in the deposition atmosphere.
Second, there is a huge {\em increase} of the RHEED intensity directly
after starting the PLD process. In contrast to the case of Sr$_2$RuO$_4$,
this effect is observed reproducibly after each growth stop and not only if
the deposition is started on the bare substrate. Note that the RHEED
intensity vs. time curves in Fig.~\ref{fe3o4} are obtained from the $(0,2)$
diffraction spot. For this spot the electrons reflected from different
growth planes interfere constructively (in-Bragg condition). In this case
for purely coherent scattering no RHEED oscillations are expected. The fact
that we do observe RHEED oscillations is caused by multiple and diffuse,
incoherent scattering which results in an increasing (decreasing)
intensity with increasing (decreasing) step density \cite{Korte:97}.
Therefore, starting from a smooth surface the RHEED intensity first
increases due to the increasing step density. That is, a $\pi$-phase shift
of the intensity oscillation relative to those recorded for off-Bragg
condition is obtained. A similar behavior has also been found for
semiconductor growth \cite{Dobson:87,Joyce:88,Braun:99}. This
interpretation is consistent with a third observation: after stopping the
growth of Fe$_3$O$_4$ the intensity {\em de\/}creases. Since it is natural
to assume that annealing leads to a smoother surface, i.e. smaller step
density, it is evident that the measured intensity of the $(0,2)$ spot is
decreasing with time. The relevance of multiple scattering processes is
also manifested by the presence of Kikuchi lines. The RHEED pattern in the
inset of Fig.~\ref{fe3o4} indicates that the resonant enhancement is due
to both, coupling to surface bound states, and three-dimensional
diffraction \cite{Braun:99}.

From the measured film thickness and the number of RHEED oscillations, for
Fe$_3$O$_4$ we obtain {\em four} oscillations per unit cell. Considering
the unit cell of Fe$_3$O$_4$ with the inverse spinel Fd$\overline{3}$m
structure ($a=b=c=0.8396$\,nm), one can identify the corresponding
molecular layers each consisting of the composition
Fe(A)$_2^{3+}$Fe(B)$_2^{3+}$Fe(B)$_2^{2+}$O$_8^{2-}$. The letters A and B
refer to the tetrahedral A-site and the octahedral B-site, where the B-site
is equally occupied by Fe$^{3+}$ and Fe$^{2+}$ ions. These building blocks
cannot be mapped onto each other by lattice translations. We note that our
high quality, 30-50\,nm thick magnetite films have a saturation
magnetization at room temperature close to the theoretically expected value
of 4.0~$\mu_{B}$/f.u., that is, the saturation magnetization is comparable
to the best results reported in literature so far \cite{Kale:01}. We
further note, that recently four RHEED intensity oscillations per unit cell
have been observed for the layered manganites. Interestingly, there the
basic building block corresponds to only half the layer thickness expected
according the chemical composition $\frac{1}{2}\cdot$(La,Sr)$_3$Mn$_2$O$_7$
\cite{Philipp:02}.


In summary, we have grown Sr$_2$RuO$_4$, MgO, and Fe$_3$O$_4$ epitaxial
thin films by PLD from stoichiometric targets. Our results show that high
pressure RHEED allows to monitor the epitaxial growth of the complex oxide
materials on a sub-unit cell level.  We have observed a molecular or block
layer growth mode where the basic building blocks are determined by the
chemical composition provided by the stoichiometric target material. For
materials with units cells consisting of several block layers of identical
stoichiometry, several RHEED oscillations per unit cell are observed.

This work was supported in part by the Deutsche Forschungsgemeinschaft
(project Al/560) and the BMBF (project 13N8279). We thank S.~Schymon for
the preparation and characterization of Sr$_2$RuO$_4$ thin films.


\end{document}